\documentclass[conference]{IEEEtran}
\usepackage[dvips]{graphicx}
\usepackage{cite}
\usepackage{psfrag}
\usepackage{subfigure}
\usepackage{theorem}

\usepackage{enumerate}
\usepackage{times}
\usepackage{amsmath}
\usepackage{comment}
\DeclareMathOperator*{\argmax}{arg\,max}
\DeclareMathOperator*{\argmin}{arg\,min}

\begin{document}
\title{Reinforcement Learning Based Transmission Strategy of Cognitive User in IEEE 802.11 based Networks}


\author{
\authorblockN{Rukhsana Ruby\authorrefmark{1}, Victor C.M. Leung\authorrefmark{1}, John 
Sydor\authorrefmark{2}}
\authorblockA{\authorrefmark{1}Department of Electrical and Computer Engineering,\\
The University of British Columbia,\\
Vancouver, BC, Canada}
\authorblockA{\authorrefmark{2}Communication Research Centre (CRC),\\
Ottawa, Ontario, Canada}
}

\maketitle

\begin{abstract}
Traditional concept of cognitive radio is the coexistence of primary and secondary user in multiplexed manner. we consider the opportunistic channel access scheme in IEEE 802.11 based networks subject to the interference mitigation scenario. According to the protocol rule and due to the constraint of message passing, secondary user is unaware of the exact state of the primary user. In this paper, we have proposed an online algorithm for the secondary which assist determining a backoff counter or the decision of being idle for utilizing the time/frequency slot unoccupied by the primary user. Proposed algorithm is based on conventional reinforcement learning technique namely Q-Learning. Simulation has been conducted in order to prove the strength of this algorithm and also results have been compared with our contemporary solution of this problem where secondary user is aware of some states of primary user.
\end{abstract}


\section{Introduction}
In terms of the role in increasing the efficiency
and aggregate network throughput, cognitive radio concept plays differently than the conventional spectrum
allocation methods~\cite{Qing07, Simeone07, Hang08, Haykin05}. In cognitive networks, unlicensed
secondary users opportunistically access radio bandwidth
owned by licensed primary users in order to maximize their
performance, while limiting interference to primary users’
communications.

Previously, cognitive radio mostly focused on a white space approach~\cite{Simeone07},
where the secondary users are allowed to access only those
time/frequency slots left unused by the licensed users. White space approach is based on zero interference rationale. But, due to noise and fading in channel and mechanism of channel sensing, errors in measurement are inevitable~\cite{Yunxia08}. Therefore, in practical scenarios,
there is some probability of having collision between primary and secondary users, which can be measured and used as a constraint
for the optimization problem. There are some works investigating
the coexistence of primary/secondary signals in the same
time/frequency band by focusing on physical layer methods for
static scenarios, e.g.,~\cite{Haykin05, Wenyi10, Lili08, Hang08, Yiping07}. Considering the dynamism while superimposition of primary and secondary users on the same time/frequency slot, a strategy of secondary user has been derived where the primary user operates in slotted ARQ based networks~\cite{Levorato09, Levorato092}.

We consider IEEE 802.11 based networks where primary users follow DCF protocol in order to access the channel. Unlike the work~\cite{Levorato09}, in our contemporary work~\cite{RRuby11}, we have developed a transmission strategy for the secondary user which picks a backoff counter intelligently or remains idle after having a transmission in a multiplexed manner. As the user needs to pass DIFS and backoff time period before flushing a packet into the air, the secondary user does not know the exact state of the primary user. Therefore, the performace constraint of the primary user plays a great role in the decision making process of secondary user. Our previous work revealed solution by formulating the problem as linear program being assumed that secondary user does know the traffic arrival distribution of primary user.

As this approach assumes that the secondary transmitter has some knowledge of the current state and probabilistic model of the primary transmitter/receiver pair, limiting its applicability. For example, while it is likely that the secondary might read ACKs for the primary system, it is unlikely that the secondary will have knowledge of the pending workload of packets at the primary transmitter or will know the distribution of packet arrivals at the primary transmitter. Therefore, we address this limitation by developing an on line learning approach that uses one feedback bit sent by the primary user and that approximately converges to the optimal secondary control policy.  We will show that when the secondary user has access to such tiny knowledge, an online algorithm can obtain performance similar to an offline algorithm with some state information.

Rest of the paper is organized as follows, section~\ref{sec:sysmodel} illustrates system model of the network, which includes the detailed optimization problem and solution thereafter. Results obtained from simulation have been shown in section~\ref{sec:perfeval} in order to verify the efficacy of the algorithm. Finally section~\ref{sec:concl} concludes the paper.

\section{System Model}
\label{sec:sysmodel}
\begin{figure}
  \begin{center}
    \includegraphics[width=0.8\columnwidth]{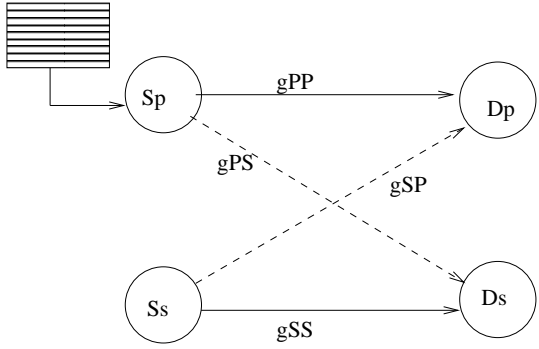}
    \caption{System Model}
    \label{fig:sysmodel}
  \end{center}
\end{figure}

\begin{figure}
  \begin{center}
    \includegraphics[width=0.8\columnwidth]{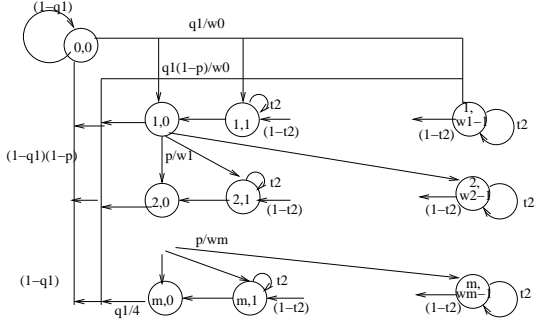}
    \caption{Markov Model of Primary User}
    \label{fig:primmarkov}
  \end{center}
\end{figure}

\begin{figure}
  \begin{center}
    \includegraphics[width=0.8\columnwidth]{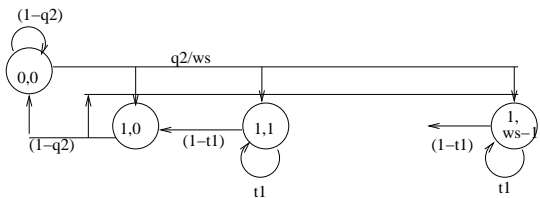}
    \caption{Markov Model of Secondary User}
    \label{fig:secmarkov}
  \end{center}
\end{figure}

We consider interference mitigation scenario in IEEE 802.11 based networks. The prime assumption on the interference mitigation strategy is that both users can decode their packets with some probability when they transmit together or individually. However, secondary user is constrained to cause no more than a fixed maximum degradation of the primary’s performance. This approach is the other end of white space one. If primary user cannot tolerate any loss, the optimal strategy for the secondary user is not to transmit at all. whereas in the work~\cite{Levorato09}, secondary user can detect the slot occupancy and can only transmit in the slots which it finds empty and therefore incurs some throughput even if primary user cannot tolerate any throughput loss. Consider the network in figure~\ref{fig:sysmodel} with a primary and secondary source, namely $S_P$ and $S_S$. Destination of these source nodes are $D_P$ and $D_S$ respectively.

We assume a quasi static channel, and time is divided into slots. Before initiating a packet transmission, both users first undergo DIFS period and decrements the backoff counter which is as large as each single time slot. While decrementing backoff counter, if the station detects a busy channel, it halts its decrementing process and resumes until it detects idle channel for the length of DIFS period. When the counter reaches to zero, packet is flushed out into the air. Packets have a fixed size of L-bits, and transmission of a packet plus its associated feedback message fits the duration of a slot. Ideally, packet transmission time is variable, but in this work for the sake simplicity, it is constant i.e. multiple of some slots. We denote by $g_{PP}$, $g_{PS}$, $g_{SS}$ and $g_{SP}$, the random variables corresponding to the channel coefficients respectively between $S_P$ and $D_P$; $S_P$ and $D_S$; $S_S$ and $D_S$; $S_S$ and $D_P$ with $\zeta_{PP}(g)$, $\zeta_{PS}(g)$, $\zeta_{SS}(g)$ and $\zeta_{SP}(g)$ their respective probability distribution.
The average decoding failure probability at the primary destination $D_P$ associated with a silent secondary source is denoted by $\rho > 0$, while the same probability when the secondary source transmits is $\rho^* > \rho$. Analogously, the average decoding failure probability at the secondary destination $D_S$ when the primary source is silent and transmitting is denoted with $\nu > 0$ and $\nu^* > \nu$ respectively.

Control Protocols implemented by the primary user is greatly impacted by the secondary user's transmission as discussed in the above paragraph. Thus, it degrades the primary user's performance and this manner is true for the secondary user as well. However, the goal of the system design is to optimize secondary user's performance without doing harm to the primary user in some extent. Therefore, upon receiving the feedback from the primary user, secondary one adjusts its transmission policy. Packet arrival at the primary user is designed as a poisson arrival process with the parameter $\lambda_1$.

The state of the network can be modeled as a homogeneous markov process. Two parameters (backoff stage, counter value) referred to as (b, c) describe the state of a user, where $c$ can take any value between 0 and $w_b - 1$. Backoff stage b varies from 1 to maximum backoff stage $m$. Here, $m$ is the maximum retry limit. Having a transmission failure, each packet is attempted by the primary user for retransmission at most $m$ times. At each backoff stage, if a station reaches state $(b, 0)$ (i.e. backoff counter value becomes 0), the station will send out a packet. If the transmission failure occurs at this point with some probability, the primary user moves to higher backoff stage $(b+1, c)$ with probability $\frac{1}{w_{b+1}}$. If successful packet transmission happens, the primary user goes to idle state $(0, 0)$ (if there is no outstanding packet in queue) or in the initial backoff stage having picked some backoff counter with the probability of $\frac{1}{w_1}$. Markov chain model of primary user has been illustrated in figure~\ref{fig:primmarkov}. Secondary user tries each packet only once, after having transmission, it goes to idle state with some probability or picks a backoff counter $j$ with probability $\kappa(1, j+1)$ for the transmission of new packet from the queue. Note that, secondary user's packet is assumed as backlogged or there is always one packet in the queue. However, in order to meet the performance loss constraint of primary user, secondary user needs to keep silent and therefore we have introduced a fake variable $\lambda_2$ i.e. secondary user's packet arrival rate. Markov chain model for the secondary user has been shown in figure~\ref{fig:secmarkov}. In both figures, $q_1$ and $q_2$  are function of $\lambda_1$ and $\lambda_2$. Detailed state transitions and steady state distribution of the problem have been skipped in this work due to space constraint. Goal of this work to find a optimal strategy for the secondary limiting the performance loss of primary user.

\subsection{Optimization Problem}
\label{subsec:opt-prob}
Let us define the cost functions $X_i\left(\phi, u\right): {\chi}\times{\bigcup}~{\rightarrow}~R$ as the average cost incurred by the markov process in state $\phi\in\chi$ if action $u\in\bigcup$ is chosen. Note that, $u = 0$ represents the secondary source keeps silent and $u = 1$ represents the picking of a backoff counter from secondary backoff counter window i.e. $\left[0, 1, \cdots, w_s-1\right]$. And, average generic cost function yields to

\[
X_i(u)~=~\displaystyle\lim_{n->\infty}\frac{1}{n}\displaystyle\sum_{t=1}^{n}E\left[X_i(\phi_t, u_t, \epsilon_t(\phi_t, u_t))\right]
\]

where $\bigcup = \{u_1, u_2, \cdots\}$ is the sequence of actions of the secondary source and $\epsilon_t(\phi_t, u_t)$ is an exogenous random variable which is not instantaneously obtained due to protocol specific behavior. For example, if secondary user picks a backoff counter $j$, it has to go through first DIFS and $j$ times backoff slots before having transmission. While passing through the backoff slots, it might be halted by the transmission of primary user and reduces the overall throughput than the case of not being halted. This incidence is also true for the primary user as well. Moreover, state variable $\phi_t$ is not explicit to the secondary user, because secondary user does not know if the primary user is in backoff stage or in idle slot. However, secondary user can sense the primary user's presence if the primary user transmits in a slot. Considering all these issues, our high level cost functions have been derived below.


\[
\begin{array}{l}
X_0(\phi, u, \epsilon)~=~  \\
\{\begin{array}{ll}
 \theta_S(\nu) & \mbox{if}~u = 1~\&~\phi~\neq~(b, 0) \\
 \theta_S({\nu}^*) & \mbox{if}~u = 1~\&~\phi~=~(b, 0) \\
 \theta_S & \mbox{if}~u = 0~\&~\forall{\phi\in\chi} \\
\end{array} \\
\\
X_1(\phi, u, \epsilon)~=~ \\
\{\begin{array}{ll}
 \theta_P(\rho) & \mbox{if}~u = 0~\&~\phi~\neq~(0, 0) \\
 \theta_P({\rho}^*) & \mbox{if}~u = 1~\&~\phi~\neq~(0, 0)
\end{array}
\end{array}
\]

$\theta_S(\nu)$ and $\theta_S(\nu^*)$ are the instantaneous calculated secondary user's throughput assuming the failure probability of transmitted packet is $\nu$ and $\nu^*$ respectively. As discussed previously, $\nu^*$ and $\nu$ are the failure probability of secondary user's transmitted packet when primary user transmits and does not transmit respectively. Besides these two cases, $X_0$ is just the throughput of secondary user considering the current time slot as we know secondary user's queue is backlogged. Sitting idle in other's transmission time and backoff slots are taken account into the calculation of throughput.

\[
\begin{array}{ll}
X_2(\phi, u, \epsilon)~=~ & X_3(\phi, u, \epsilon)~=~ \\
\{\begin{array}{ll}
 1 & if~tx~fails~\&~\phi~=~(m, 0) \\
 0 & otherwise
\end{array} 
 &
\{\begin{array}{ll}
 1 & if~\phi~=~(1, 0) \\
 0 & otherwise
\end{array} 
\end{array}
\]

And again, $X_2(u)$ can be interpreted as the fraction of time slots in
which the primary source fails the last allowed transmission
and the packet would not be delivered and $X_3(u)$ is the
fraction of time slots where the primary begins the service of
a new packet. In this paper we define the failure probability
as the average ratio of dropped packets after $m$ retransmission,
to the total number of new packets sent, one can see that
$\frac{X_2(u)}{X_3(u)}$ is equivalent to the failure probability of the
primary source’s packets. The optimization problem is then given by

\begin{equation}
\label{eq:opt-prob}
\argmin_{\kappa}X_0(u)~~~s.t.~~\theta_P^{max}-\theta_P~\leq~\gamma_1 ~OR~\frac{X_2(u)}{X_3(u)}~\leq~\gamma_2
\end{equation}

It is shown in~\cite{RRuby11} that the optimization problem in equation~\ref{eq:opt-prob}
is solved by formulating the problem as a linear program. Parameters in the formulation have been derived from the steady state distribution of the markov chain. Finally, the obtained optimal strategy has been denoted by a vector $\kappa$. Elements of this vector holds the proportion of time secondary user keeps silent or in which probability should it picks the backoff counter from the given contention window. Solution needs a little brute force search with some standard policy that have been proven analytically.

\subsection{State Knowledge}
The offline solution of the optimization problem requires
full knowledge of state $\phi_t$, which corresponds to the transmission index and
queue state of the primary source, as well as knowledge of the
transition probabilities and cost functions. However, the full
knowledge of $\phi_t$ requires an explicit exchange of information.

We address this limitation in two steps. First, by assuming
that the secondary only has information about what can be
directly observed about the primary, and second, by using an
on line learning technique that learns the necessary parameters
without requiring knowledge of the transition probabilities.

By sensing the channel, primary user cannot instantaneously detect the channel condition as primary user follows DCF protocol. Therefore, there is no way to get the information about the state of the primary user when it is in backoff state or in idle state i.e. primary user's queue is empty. Secondary user can get to know if primary user transmits in a certain slot by sensing the channel. In some cases, secondary user can get knowledge if primary user's transmitted packet is new or old. The header includes the sequence number of the packet, which increases if the transmitted packet is a new one and remains the same if it is a retransmission. However, when the retransmitted packet reaches to its maximum limit, there is no way for the secondary user to know, in the next slot whether it will go through another backoff stage with fresh packet, since the buffer state is completely unknown to the secondary user.

Even though primary user can gather such little information, in the proposed solution, secondary user does not rely on these information. Rather, the proposed solution depends on a simple bit which indicates whether the performance constraint of the primary user is satisfied in the current time slot or not. This information is sent by the primary user as piggy backed form in either ACK or the actual packet's header. Having this information, secondary user regulates its transmission strategy.

It is shown in the following section via numerical results
that this such partial knowledge is sufficient to implement a learning
algorithm operating close to the limit provided by full state
knowledge. Note that, the state of the primary user is overlooked here, it does not help in the decision making process of the secondary user. Rather the cost functions are most important driving factor of the proposed online algorithm.

\subsection{Learning Algorithm}

Most approaches to optimal control require knowledge of an
underlying probabilistic model of the system dynamics which
requires certain assumptions to be made, and this entails a
separate estimation step to estimate the parameters of the
model. In particular, in our optimization paradigm~\cite{RRuby11}, the
optimal randomized stationary policy can be found if the
failure probabilities $\rho$, $\rho^*$, $\nu$, $\nu^*$ are known to the secondary
user, together with some knowledge of state $\phi$. In
this section we describe how we can use an adaptive learning
algorithm called Q-learning~\cite{Richard98, Mahadevan96} to find the optimal policy without
a priori knowledge about our probabilistic model.

The Q-learning algorithm is a long-term average reward
reinforcement learning technique. It works by learning an
action-value function $R_t(\phi, u)$ that gives the expected utility
of taking a given action $u$ in a given state $\phi$ and following a
fixed policy thereafter. Intuitively, the Q-function captures the
relative cost of the choice of a particular allocation for the next
time-step at a given state, assuming that an optimal policy is
used for all future time steps. Q-learning is based on the adaptive iterative learning of Q
factors. However, as discussed previously, it is almost impossible to get to know about the information of primary user's current state  and thus it ignores the current state $\phi_t$ while learning the system and behavioral parameters of primary user. Since, secondary user overlooks current state $\phi_t$ while taking any action, we can call it as the variant of markov decision process(MDP). The original MDP means, the agent takes action based on the current state of the environment.

No matter, secondary user follows MDP or variant of MDP, it needs to fix a cost function which is typically named as reward. Ultimate reward of the secondary user is its own throughput i.e. $X_0(\phi, u, \epsilon(\phi, u))$ which it wants to maximize. However, in order to maximize throughput, we have adopted some indirect approach to get the maximized value of $X_0(\phi, u, \epsilon(\phi, u))$. Cost function is associated with the action of secondary user. Proability of each action of secondary user is resided in the vector $\kappa~=~[0, 1, \dots, w_s]$. Length of this vector is $w_s + 1$ ($w_s$ is backoff window size of secondary user). Index $0$ denotes the proportion of time secondary user keeps itself silent, subsequent indexes $i$ denote the portion of time backoff counter $(i-1)$ is chosen by the secondary user. As discussed previously, outcome of secondary user's action is not obtained instantaneously until the secondary user has its transmission. Due to the interaction of secondary and primary user, the obtained throughput from each action vary and our cost function is the obtained average throughput (added to the long term average throughput) resultant from the taken action. Let $X_0^p(\phi, u, \epsilon(\phi, u))$ is the average throughput of secondary user while taking the action $u$ and $X_0^n(\phi, u, \epsilon(\phi, u))$ is the average throughput when the secondary user really completes its packet transmission. Then, the cost function at time $t$ is defined as follows:

\[
c(\phi, u, \epsilon(\phi, u))~=~X_0^n(\phi, u, \epsilon(\phi, u))~-~X_0^p(\phi, u, \epsilon(\phi, u))
\]

And our optimization problem thus stands to

\begin{equation}
\label{eq:opt}
\argmax_{\kappa}\displaystyle\lim_{n->\infty}\frac{1}{n}\displaystyle\sum_{t=1}^{n}E\left[c_t(\phi_t, u_t, \epsilon_t(\phi_t, u_t))\right]
\end{equation}

And the Q-Learning algorithm for solving equation~\ref{eq:opt} is illustrated as follows.

\begin{itemize}

\item {\bf Step 1:} Let the time step $t=0$. Initialize each element of reward vector $R_t(\kappa)$ as some small number, such as 0.

\item {\bf Step 2:} Check if the constraint of the primary user is satisfied. If not, choose the action of $u = 0$. Otherwise, choose the action $u$ with index $j = [0, 1, \cdots, w_s]$ that has the highest $R_t(j)$ value with some probability say $1 - \tau_t$, else let $u$ be a random exploratory action. In other words,

\[
u_t~=~\argmax_{\kappa}R_t(\kappa)
\]

\item {\bf Step 3:} Carry out action $u_t$. Wait until secondary user completes its transmission if it picks any backoff counter. Or secondary user may choose the option of being silent. In either case, Calculate the cost function $c_t(u_t)$ and update the reward variable for the corresponding action. If the current state is $\phi$ and the resultant state is $\phi'$ after taking the action  $u_t$, reward is updated as follows.

\begin{equation}
\label{eq:q-learning}
R_t(\phi, u_t)~=~(1~-~\alpha_t)R_t(\phi, u_t) + \alpha_t(c_t + {\gamma}\displaystyle\max_{u'}R_t(\phi', u'_t))
\end{equation}

\item {\bf Step 4:} Set the current state as $\phi'$ and repeat step 2. When convergence is achieved, set $\tau_t = 0$.

\end{itemize}

This is the typical Q-learning algorithm. In our case, we don't know the primary user's exact current state and also don't know what the next state will be. Therefore the equation~\ref{eq:q-learning} reduces to

\[
R_t(u_t)~=~(1~-~\alpha_t)R_t(u_t) + {\alpha_t}c_t
\]

In order to obtain the optimal value of $\alpha_t$, we have found the following theorem.

{\bf Theorem 1:} Step size parameter $\alpha_t = \frac{1}{t}$ gives the convergence to the algorithm.

{\bf Proof:} The choice $\alpha_t = \frac{1}{t}$ results in the sample-average method, which is guaranteed to converge to the true action values by the law of large numbers. A well-known result in stochastic approximation theory gives us the conditions required to assure convergence with probability 1:

\[
\begin{array}{lll}
  \displaystyle\sum_{t=1}^{\infty}\alpha_t~=~\infty
 &
 and &
  \displaystyle\sum_{t=1}^{\infty}{\alpha}_t^2~<~\infty
 \\
\end{array}
\]

The first condition is required to guarantee that the steps are large enough to eventually overcome any initial conditions or random fluctuations. The second condition guarantees that eventually the steps become small enough to assure convergence. Note that, both convergence conditions are met for the sample-average case, $\alpha_t = \frac{1}{t}$.

\section{Performance Evaluation}
\label{sec:perfeval}
In this section we will evaluate the performace of our online algorithm. In addition, we have compared performance of this algorithm with the algorithm which has some information of primary user as presented in our work~\cite{RRuby11}. Throughout the simulation, we assume that the buffer size of
the primary source is $B = 4$ and the maximum retransmission
time is $m = 4$. Backoff window size in each stage is 4, 6, 8, and 10 respectively. Secondary user's backoff window size is as $w_s = 3$. We set the failure probabilities for the transmission of the primary source $\rho~=~0.2$, $\rho^*~=~0.5$, depending on
the fact that secondary is silent or not, respectively. Similarly,
the failure probabilities of the secondary source are set to be
$\nu~=~\nu^*~=0.3$. Note that these failure probabilities
are not known at the secondary source and it has to learn the
optimal policy without any assumption on these parameters in
advance. Once again, the goal of the algorithm is to maximize the throughput of the secondary source.

\begin{figure}
  \begin{center}
    \includegraphics[width=0.8\columnwidth]{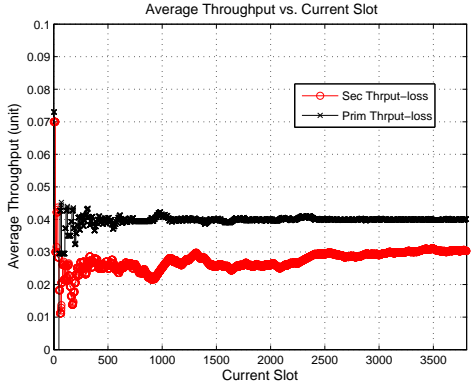}
    \caption{Average Throughput (unit) vs. Current Slot for $\lambda_1 = 0.05$ and $\gamma_1 = 0.04$}
    \label{fig:thrput-convergence}
  \end{center}
\end{figure}

Figure~\ref{fig:thrput-convergence} depicts the convergence of secondary and primary user's throughput from 0'th iteration to some number of iterations. Throughput loss is defined as the difference between maximum achievable throughput and instantaneous throughput at a particular slot. From the given parameters, maximum achievable throughput is calculated considering only a single user (primary or secondary) is acting on the channel. We see the convergence of throughput loss happens after a few iterations.

\begin{figure}
  \begin{center}
    \includegraphics[width=0.8\columnwidth]{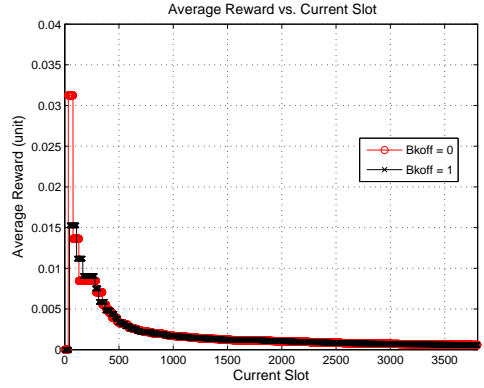}
    \caption{Secondary User's Average Reward (unit) vs. Current Slot for $\lambda_1 = 0.05$ and $\gamma_1 = 0.04$}
    \label{fig:reward-convergence}
  \end{center}
\end{figure}

In order to extrapolate the cost functions of our algorithm, we also have shown convergence process of two actions picked up by the secondary user, i.e. probability of picking backoff counter 0 and 1 respectively in the figure~\ref{fig:reward-convergence}. We have initialzed cost of all actions at time slot zero. As the algorithm moves along with time, it updates its average reward according the formula presented in the algorithm. The algorithm is more prone to pick backoff counter with lower value that will be shown in the subsequent figures. However, in terms of general rule, algorithm does not pick the same action repeatedly. This is because, due to the interaction between primary and secondary users, the repeated action may cause to the degradation of the primary user's performance or it may degrade its own average reward value than the other actions. Consequently, the algorithm moves to the other action and the average reward value over the time for different actions look similar.

\begin{figure}
  \begin{center}
    \includegraphics[width=0.8\columnwidth]{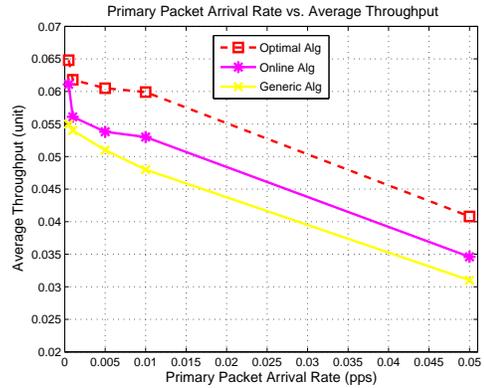}
    \caption{Primary User's Packet Arrival Rate (pps) vs. Average Throughput for $\gamma_1~=~0.04$}
    \label{fig:secthrput-comparison}
  \end{center}
\end{figure}

Figure~\ref{fig:secthrput-comparison} shows the throughput of primary and secondary source with the increased packet arrival rate $\lambda_1$ for a fixed tolerable primary source's throughput loss. As expected, throughput of the secondary source decreases as $\lambda_1$ is increased gradually. A larger $\lambda_1$ means that the primary source is accessing the channel more often. Therefore, the number of slots in which the secondary source can transmit while meeting the constraint on the throughput loss of the primary source decreases. In addition, in this figure, we have projected the result obtained by our optimal algorithm~\cite{RRuby11}. Optimal algorithm though due to the protocol behavior is not fully aware of state of the system, has some better information than our proposed online algorithm. Therefore, it incurs better performance in terms of achievable throughput for different $\lambda_1$ value. Whereas, our online algorithm though does not look like have similar performance, but gains better one than other blind generic algorithm. Generic algorithm means, here secondary user picks its backoff counter uniformly. With this strategy, we see the performance for the secondary user is the worst. Even worst news is that, this algorithm is completely blind about the performance constraint of primary user.

\begin{figure}
  \begin{center}
    \includegraphics[width=0.8\columnwidth]{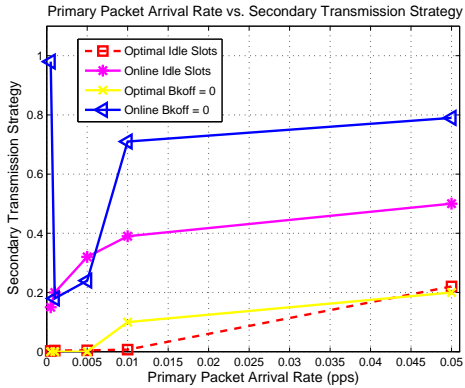}
    \caption{Primary User's Packet Arrival Rate (pps) vs. Secondary Transmission Strategy for $\gamma_1~=~0.04$}
    \label{fig:secstrategy-comparison}
  \end{center}
\end{figure}

Figure~\ref{fig:secstrategy-comparison} compares the obtained secondary user's strategy for both our optimal and online algorithms. We have presented the proportion of idle slots and probability of picking backoff counter 0. For the sake of page limit, we have skipped other results here. In this result apparently, we don't see any match between two algorithms. However, we can explain the difference. In fact, online algorithm is mostly dependent on the primary user' performance loss violation indicator and its own reward value for different actions. It tries to pick the action with maximum value, which is usually the backoff counter with lower value. Otherwise, upon the signal of constraint violation, it keeps silent. Therefore, we see that online algorithm puts more weights to the backoff of lower value and again backoff counter of lower value breaks the constraint more often and thus it keeps more silent than offline algorithm. Whereas, optimal algorithm knows the arrival rate of primary user, it runs a near brute-force algorithm in order to find the optimal strategy of secondary user.

\section{Conclusion}
\label{sec:concl}
We have proposed an on line learning approach in interference
mitigation adopted IEEE 802.11 based networks for the cognitive user. Our approach relies only
on the little performance violation feedback of the primary transmitter and uses
Q-learning to converge to nearly optimal secondary transmitter
control policies. Numerical simulations suggest that this approach
offers performance that is close to the performance of the
system when complete system state information is known. Although, the strategy of both algorithms does not follow the exactly similar trend.

\end{document}